\documentclass[11pt]{article}
\usepackage[top = 2 cm, bottom = 2.3 cm, left = 2.5 cm, right = 1.8 cm]{geometry}
\usepackage{authblk, cite, color, amssymb, amsmath, enumitem}
\usepackage[hypertex, colorlinks = true, linkcolor = blue, citecolor = red]{hyperref}
\usepackage[dvipsnames]{xcolor}

\begin{document}

\title{Einstein's hole argument and Schwarzschild singularities}

\author[1,2]{Merab Gogberashvili \thanks{gogber@gmail.com}}
\affil[1]{Javakhishvili State University, 3 Chavchavadze Ave., Tbilisi 0179, Georgia}
\affil[2]{Andronikashvili Institute of Physics, 6 Tamarashvili St., Tbilisi 0177, Georgia}
\maketitle

\begin{abstract}
According to the Einstein hole argument, vacuum metric solutions are equivalent only if they correspond to the same energy--momentum tensor in the source region. In this paper it is shown that singular coordinates that are used to show Schwarzschild geodesics completeness, introduce the fictive delta-like sources at the horizon. Then, metric tensors obtained by such singular transformations, cannot be considered as solutions of the same Einstein equations with the central source.
\end{abstract}

%%%%%%%%%%%%%%%%%%%%%%%%%%%%%%%%%%%%%%%%%%%%%%%%%%%%%%%%%%%%%%%%%%%%%%%%%%

\section{Introduction}

Despite being the most successful theory in explaining gravitational interaction, General Relativity (GR) has several problems: In the observational side it fails to adequately address the dark sector (see the recent review \cite{Oks:2021hef}), while as the theoretical unresolved issues one can mention the information and singularity problems (see the recent reviews \cite{Marolf:2017jkr, Witten:2019qhl}). The singularity theorems and initial value problem in GR lead to the restriction that admissible coordinate transformations be of class $C^2$ \cite{Man-San, Sorkin:1995ca, Synge}, otherwise they will result in fictitious extra sources in the Einstein equations. Indeed, the class $C^0$ coordinate transformations introduce the $\delta$-like terms in the Riemann tensor, even if the initial metric tensor was smooth. This means that untransformed and transformed metric tensors are solutions of different equations and do not coincide on the surface of discontinuity.

Coordinate transformations in GR can be restricted also using so-called Einstein's hole argument \cite{hole}. In 1915, Einstein posited a specific thought experiment in which an empty spacetime domain (hole) is surrounded by a source region with non-zero energy-momentum ($T_{\mu\nu} \ne 0$). He argued that an active diffeomorphism, which acts as the identity in the source region, but is not an identity in the hole, would modify the metric field in the hole. The naive conclusion is that the energy-momentum sources cannot uniquely determine the metric in the hole. This apparent paradox, called 'the Einstein hole argument', is related to a gauge freedom in GR. It shows that the viewpoint of spacetime substantivalism, which considers geometry independently of the physical processes, leads to unpalatable conclusions in a large class of models. The metric is not a prescribed property of a Riemannian manifold, each solution of Einstein equations determines coordinates in terms of which it is written and transformed metrics may give different physical meanings to the same set of local coordinates.

For usual field equations, like Maxwell's system, the source and boundary conditions determine the field everywhere. However, the equations of classical electrodynamics do not determine the vector potential, since it depends on an arbitrary choice of gauge. Similarly, Einstein's field equation is covariant and the metric tensor can be transformed arbitrarily in active diffeomorphisms. Therefore, to avoid the multiplicity of solutions, one should assert that all the metric solutions are physically equivalent and deny that local coordinates represent real geometry \cite{Act-Pas}.

For the case of a Black Hole (BH), according to the Einstein hole argument, outer vacuum metric solutions are equivalent only if they correspond to the same under-horizon sources. Even though BHs are classical solutions of GR, they can be also understood as macroscopic quantum objects. Inside a BH it is expected that quantum effects of gravity are dominated and a regular description of the spacetime is possible in some quantum sense, analogous to the case of the Coulomb potential with regular wave functions at the central singularity. Observationally nothing is known about the interior of a BH \cite{Cardoso:2019rvt}, and it might be possible that BHs without a singularity emerge within a classical theory. A large number of  physically reasonable non-singular BH models, curvature invariants of which are finite for all points in space time, have been proposed \cite{Ansoldi:2008jw, Cardoso:2019rvt}.

The inconsistency of GR with the principles of quantum mechanics above the Planck scale motivates the need to modify the classical description of BHs not only at central singularity, but already at horizon scales \cite{Modif-1, Modif-3}. BHs can be considered as horizon size massive spheres, since they are objects formed by gravitational collapse that shrink matter to a minimum size that does not violate quantum theory \cite{Malafarina:2017csn}. This picture reconciles  two controversial views on the BH horizon: one is that the energy density there is small \cite{Howard:1984qp}, and the other is the existence of a structure like 'brickwall' \cite{tHooft:1984kcu}, 'fuzzball' \cite{Mathur:2005zp}, or 'firewall' \cite{Modif-2}, which have a large back-reaction to the geometry. Note that, as many physicists today, Einstein and Schwarzschild also believed that the event horizon is a physical singularity and matter cannot reach it \cite{Einstein, Schwarz}.

In this paper aspects of the problems with singular coordinate transformations for the Schwarzschild metric are considered, generalizations for the more realistic Kerr spacetime are straightforward. It is shown that the Einstein hole argument restricts classes of singular coordinate transformations that are usually       used to restore geodesics completeness at the Schwarzschild horizon.

The paper is organized as follows. In Sec.~\ref{Sch-BH} problems of realization of so-called Schwarzschild BHs are discussed. In Sec.~\ref{Hole} the Einstein hole arguments for the Schwarzschild solution are canalized. Vacuum spherically symmetric metrics, written in different coordinates, can be considered as equivalent only if they correspond to the same energy-momentum tensor in the under-horizon source region. In Sec.~\ref{Regge-Wheeler} it is shown that the Regge-Wheeler 'tortoise' radial variable, and corresponding singular coordinates used to demonstrate geodesics completeness, are unacceptable class $C^0$ functions and lead to the appearance of fictive sources. Due to the hole argument, such singular transformations are unacceptable. In Sec.~\ref{Geodesic} problems of singular lagrangians for geodesics are considered. The existence of solutions of particle equations that contain the exponentially time-dependent factors (with the complex phases) is justified. The Sec.~\ref{Concl} is devoted to the concluding remarks. It is argued that particles probably do not cross the BH event horizon, but are absorbed or reflected by it.

%%%%%%%%%%%%%%%%%%%%%%%%%%%%%%%%%%%%%%%%%%%%%%%%%%%%%%%%%%%%%%%%%%%%%%%

\section{Schwarzschild's black holes} \label{Sch-BH}

The most important solution of GR is the Schwarzschild metric:
\begin{equation} \label{Schwarzschild}
ds^2 = \left( 1 - \frac {r_S}{r}\right) dt^2 -  \left( 1 - \frac {r_S}{r}\right)^{-1}dr^2 - r^2 d\Omega^2~,
\end{equation}
where
\begin{equation} \label{2-sphere}
d\Omega^2 = d\theta^2 + \sin^2 \theta d\phi^2
\end{equation}
is the metric of the unit 2-sphere of the area $4\pi$ and the parameter $r_S = 2GM$ determines the Schwarzschild horizon.

The metric (\ref{Schwarzschild}) mainly is used to describe exterior geometry of a classical spherical body. To assign the value $r_S$ to an integration constant in a general spherically symmetric solution of vacuum Einstein equations, one needs to find a suitable static interior solution and introduce junction conditions on its surface. In Schwarzschild's coordinates (\ref{Schwarzschild}), the gravitational mass of a body, which determines the parameter $r_S$, is given by
\begin{equation}
M = \int 4\pi \rho (r) r^2 dr~,
\end{equation}
where $\rho (r)$ represents the density function. To the best of the knowledge, no corresponding study has ever been made for other coordinates. For a recent review about problems with the definition of BHs mass see \cite{Ha-2}.

Often (\ref{Schwarzschild}) is also used to represent so-called Schwarzschild BHs, when $r < r_S$. In this case the 'mass' $M$ is assumed not to located anywhere (since the stress-energy-momentum tensor is identically zero everywhere) and is considered just as a useful concept describing not 'fully formed' BH from the point of view of a distant observer, who sees the matter that in the process of collapsing is 'frozen' just before the horizon. In this case the metric (\ref{Schwarzschild}) is thought to represent the end state of collapse after an infinite amount of time. When the collapse is complete, all the matter that formed the BH vanishes in central singularity at $r = 0$, with zero surface area, leaving behind the vacuum gravitational field \cite{BH-1, BH-2, BH-3}.

In Schwarzschild BHs studies it is often supposed that $r$ in (\ref{Schwarzschild}) is like a proper radial coordinate, which can go down to zero. However, the parameter $r$ does not of itself determine any distance, but represents the inverse square root of the Gaussian curvature of a spherically symmetric geodesic surface in the spatial section, i.e. is the radius of curvature of a two-sphere (\ref{2-sphere}) \cite{Levi-Civita, Crothers}. Because of this, without disturbing spherical symmetry and violation of the field equations, $r$ can be replaced by any analytic function. In addition, the point at the center of spherical symmetry need not be coincident with the origin of the coordinate system.

In general, the proper distance must be calculated by the geometrical relations intrinsic to the concrete metric. In the Minkowski space-time, the proper and curvature radii coincide,
\begin{equation}
\int \sqrt{- g_{rr}} dr = \int_0^r dr = r~,
\end{equation}
and one can construct the scalar quantity $r$, representing the magnitude of the radius vector $\vec r$ from the origin point of the coordinate system (having zero volume). Then, the associated spherical surface, $4\pi r^2$, can be shrunk to zero. In any other case, when $g_{rr} \ne const$, the central sphere has some nonzero volume, i.e. the curvature scalar $r$ cannot go down to zero.

The Schwarzschild solution (\ref{Schwarzschild}) appears to introduce singularities also at $r = r_S$, entail divisions by zero or multiplication by infinity in some geometrical quantities. The singularity at $r = r_S$ is called a coordinate singularity, which can be avoided by changing to 'good' coordinates. While $r = 0$ is considered as a true physical singularity, which appears in quantities that are independent of the choice of coordinates, like the Kretschmann invariant:
\begin{equation} \label{Kretschmann}
{\cal R}^{\alpha\beta\gamma\delta}{\cal R}_{\alpha\beta\gamma\delta} = \frac {12 r_S^2}{r^6} ~.
\end{equation}
If $r \to 0$ this quantity becomes infinite, i.e. the metric cannot be extended in a smooth manner \cite{BH-1, BH-2, BH-3}. From the other hand, for sufficiently large values of BH mass, $M \sim r_S$, one can let (\ref{Kretschmann}) be arbitrarily small at $r \to r_S$. Regular features of scalar invariants at the horizon is the main motivation to use various singular radial variables (instead of $r$), which are defined on infinite or semi-infinite intervals. In these singular coordinates the Schwarzschild horizon singularity seems to 'disappear' and cannot prevent classical particles from reaching the central naked singularity at $r=0$. However, the conclusion on a finiteness of (\ref{Kretschmann}) at $r = r_S$ is usually based on an assumption of a mutual cancellations of delta-like divergences. The same is true for other invariants of the gravitational field at the horizon. For example, the determinant of (\ref{Schwarzschild}),
\begin{equation} \label{determinant}
g = \sqrt{ g_{tt}\cdot g_{rr}} ~r^2 \sin \theta  = \sqrt{ \frac {r_S - r}{r_S - r}} ~r^2 \sin \theta \quad \to \quad r^2 \sin \theta ~,
\end{equation}
is also ill-defined at $r = r_S$, and is unstable under small variations of $g_{tt}$ and $g_{rr}$. In general, $g_{tt}$ and $g_{rr}$ are independent functions and the cancellation of their zeros at the horizon is accidental, since it follows from the exact validity of the vacuum Einstein equations implying a perfect sphericity. However, a perfect spherical symmetry and true vacuums are rarely observed, if ever.

For example, one can consider a model with the time-dependent parameter $r_S$ for a collapsing BH, since any particle (even a single photon) traversing the horizon will change its mass. For a time-dependent version of the metric (\ref{Schwarzschild}) we immediately recover physical singularities in various scalars at $r \to r_S$. e.g. the Ricci scalar for time-dependent Schwarzschild mass obtains the form:
\begin{equation}
{\cal R} = \frac{r\left[ \ddot r_S(r_S - r) - 2\dot r_S^2\right]}{(r_S - r)^3}~,
\end{equation}
where overdots mean time derivatives.

Note also the equations of motion for the system of classical particles in the quadrupole approximation \cite{Dix},
\begin{equation} \label{quad}
\frac{Dp^\mu}{ds} = F^\mu = - \frac 12 {\cal R}^\mu{}_{\nu \alpha\beta}u^\nu S^{\alpha \beta} - \frac 16 J^{\alpha\beta\gamma\delta}D^\mu {\cal R}_{\alpha\beta\gamma\delta}~,
\end{equation}
where $J^{\alpha\beta\gamma\delta}$ is the quadrupole moment, $S^{\alpha \beta}$ is the spin tensor and $u^\nu$ is the 4-velocity. We see that the force (\ref{quad}) diverges at the Schwarzschild horizon, since from six non-zero components of the mixed Riemann tensor, the three
\begin{equation}
{\cal R}^t{}_{rrt} = 2{\cal R}^\theta{}_{r\theta r} = 2{\cal R}^\phi{}_{r\phi r} = \frac {r_S}{r^2(r_S - r)}~,
\end{equation}
blow up at $r = r_S$.

%%%%%%%%%%%%%%%%%%%%%%%%%%%%%%%%%%%%%%%%%%%%%%%%%%%%%%%%%

\section{Einstein's hole argument for Schwarzschild} \label{Hole}

Let us consider various aspects of the Einstein hole argument for the case of a spherical symmetric source that is concentrated under the Schwarzschild horizon, $r \le r_S$ (source region) \cite{Macdonald}. If we perform  a change of the radial coordinate in the outer metric (hole region),
\begin{equation} \label{r=f}
r = f(R)~, \qquad \left \{
\begin{array} {lr}
R = r \quad (r \leq r_S)\\
R \neq r \quad (r > r_S)
\end{array}
\right.
\end{equation}
the Schwarzschild metric (\ref{Schwarzschild}) obtains the form:
\begin{equation} \label{Schwarzschild-R}
ds^2 = \left[ 1 - \frac {r_S}{f(R)}\right] dt^2 -  \left[ 1 - \frac {r_S}{f(R)}\right]^{-1} f'^2 (R) dR^2 - f^2 (R) d\Omega^2~,
\end{equation}
where $f'$ denotes the derivative with respect to $r$. The variable $R$ in (\ref{Schwarzschild-R}) is the same as $r$ inside the matter source (at $r \leq r_S$), but is different outside (in hole). As the Einstein equations are generally covariant, the both (\ref{Schwarzschild}) and (\ref{Schwarzschild-R}) are solutions, they represent the same gravitational field on the same spacetime manifold and should be physically equivalent. For instance, in (\ref{Schwarzschild-R}) the sphere at $R$ has the area $4\pi f^2 (R) = 4\pi r^2$, as does the sphere at $r$ for (\ref{Schwarzschild}).

Now consider the metric:
\begin{equation} \label{Schwarzschild-R-r}
ds^2 = \left[ 1 - \frac {r_S}{f(r)}\right] dt^2 - \left[ 1 - \frac {r_S}{f(r)}\right]^{-1} f'^2 (r) dr^2 - f^2 (r) d\Omega^2~,
\end{equation}
which is obtained by replacing $R$ in (\ref{Schwarzschild-R}) with the Schwarzschild radial variable $r$. The metric (\ref{Schwarzschild-R-r}) has the same mathematical form as the solution (\ref{Schwarzschild-R}) and also should be a solution to the vacuum Einstein equations. The coordinate description of the interior of the source should be unaffected by the transformation, but the functional form of the Schwarzschild metric outside is changed. So, the solutions (\ref{Schwarzschild}) and (\ref{Schwarzschild-R-r}) are physically distinguishable -- in (\ref{Schwarzschild}) the sphere at $r$ has area $4\pi r^2$, while in (\ref{Schwarzschild-R-r}) at the same radius it has the area $4\pi f^2(r)$. One can conclude that a source can lead to many seemingly different metrics and Einstein equations do not uniquely determine geometry of the central mass. But, any two fields which only differ by the transformations of the type (\ref{r=f}) must be physically equivalent, similar to the case of two different vector potentials which differ by a gauge transformation.

The resolution of this seeming paradox is that the assumption of the hole argument that we have two gravitational fields (described by (\ref{Schwarzschild}) and (\ref{Schwarzschild-R-r})) on the same spacetime and using the same coordinates, is wrong. A theory that has a generally covariant field equation cannot consider two gravitational fields on the same space-time. In GR spacetime geometry and gravity are indistinguishable, we cannot start with a geometry and then introduce a gravitational field.

In order to avoid problems with the Einstein hole arguments for the Schwarzschild metric, we need to show that the meanings of $t$ and $r$ coordinates in (\ref{Schwarzschild}) and (\ref{Schwarzschild-R-r}) are different. In GR a metric defines not only the gravitational field, but also the coordinate system in which it is presented. There is no other source of information about the coordinates apart from the expression for the metric. In (\ref{Schwarzschild}) the physical meaning of $r$ follows from the $r^2 d\Omega^2$ term: events on the sphere of area $4\pi r^2$ have the radial coordinate $r$. The physical meaning of $t$ also can be read off from (\ref{Schwarzschild}): the proper time for a clock at fixed $r$ and $\Omega$ is
\begin{equation} \label{t-Schwarzschild}
d\tau^2 = \left(1 - \frac {r_S}{r}\right)dt^2~.
\end{equation}
This defines the coordinate time interval $dt$ in terms of the proper time $d\tau$ measured by the clock. The physical meaning of coordinates in (\ref{Schwarzschild-R-r}) is different from ones in (\ref{Schwarzschild}), i.e. this metric is the different solution to the Einstein equations, with another source.

It should be emphasized that, according to Einstein's hole argument, different solutions in the hole region can be considered as equivalent only if they correspond to the same energy-momentum tensor in the source domain. So, the coordinate transformation (\ref{r=f}) for outer Schwarzschild metric (in the hole region) is admissible only if it does not disturb the sources at $r \leq r_S$.

%%%%%%%%%%%%%%%%%%%%%%%%%%%%%%%%%%%%%%%%%%%%%%%%%%%%%%%%%%%%%%%%%%%

\section{Regge-Wheeler's coordinate and geodesic completeness} \label{Regge-Wheeler}

Important ingredient of singular transformations, which is often used to show geodesic completeness for the Schwarzschild metric, is the so-called Regge-Wheeler's tortoise coordinate \cite{BH-1, BH-2, BH-3}:
\begin{equation} \label{tortoise}
r^* = \int  \left( 1 - \frac {r_S}{r}\right)^{-1}dr = r + r_S \ln \left( \frac {r}{r_S} - 1\right)~. \qquad \left \{
\begin{array} {lr}
&r_S < r < \infty \\
&- \infty < r^* < \infty
\end{array}
\right.
\end{equation}
The variable (\ref{tortoise}) satisfies the equation
\begin{equation} \label{tortoise-eq}
\frac {dr^*}{dr} = \left( 1 - \frac  {r_S}{r}\right)^{-1}~,
\end{equation}
in the interval of its definition that often is assumed to include the point $r = r_S$ as well. In terms of the tortoise coordinate the Schwarzschild metric (\ref{Schwarzschild}) becomes
\begin{equation} \label{Schwarzschild-Tortoised}
ds^2 = \left[1 - \frac {r_S}{r(r^*)}\right] (dt^2 - dr^{*2}) - r^2(r^*) d \Omega^2~,
\end{equation}
where $r$ is thought of as a analytic function of $r^*$.

Analytic expressions of $r$ by $r^*$ can be found only in asymptotic regions. For the large distances from the horizon, $r \to \infty$ ($r^* \to \infty$), radial variables exhibit the behaviors,
\begin{equation} \label{r*=infty}
\begin{split}
r^*(r) = & ~ r + r_S \ln {\frac {r}{r_S}} +{\mathcal {O}}\left( \frac{r_S}{r}\right) ~, \\
r (r^*) = & ~r^* - r_S\ln {\frac {r^*}{r_S}} + {\mathcal {O}}\left( \frac{r_S}{r^*}\right) ~.
\end{split}
\end{equation}
Close to the event horizon, $r \to r_S$ ($r^* \to - \infty$), we have
\begin{equation} \label{r*=-infty}
\begin{split}
r^* (r) =& ~ r_S  \ln \left( \frac {r}{r_S} - 1 \right) + {\mathcal {O}}\left( r - r_S \right)~, \\
r (r^*) =& ~ r_S + r_S e^{r^*/r_S} + {\mathcal {O}}\left( r_S^2 e^{2 r^*/r_S}\right)~,
\end{split}
\end{equation}
and the metric (\ref{Schwarzschild-Tortoised}) obtains the following approximate form:
\begin{equation} \label{Schwarzschild-Tortoised-approx}
ds^2 \approx e^{r^*/r_S} (dt^{*2} - dr^{*2}) - r_S^2 d\Omega^2~.
\end{equation}
This metric describes meanings of coordinate parameters, $t^*$ and $r^*$, close to the horizon. From the $r_S^2 d\Omega^2$ term in (\ref{Schwarzschild-Tortoised-approx}) we see that events on the horizon (having the area $4\pi r_S^2$), can be represented by the radial coordinate $r_S$, as in (\ref{Schwarzschild}). However, for a clock at fixed $r^*$ and $\Omega$, the proper time is
\begin{equation} \label{t-tortoise}
d\tau^2 \approx e^{r^*/r_S}dt^{*2}~,
\end{equation}
which is very different from the behavior (\ref{t-Schwarzschild}). The meanings of time coordinates in (\ref{t-Schwarzschild}) and (\ref{t-tortoise}) are different, i.e. (\ref{Schwarzschild}) and (\ref{Schwarzschild-Tortoised}) are the different solutions to the Einstein equations, corresponding to different sources. Thus, due to the Einstein hole argument, the introduction of the tortoise radial variable (\ref{tortoise}) cannot be considered as a coordinate transformation of the Schwarzschild solution (\ref{Schwarzschild}).

To clarify this issue note that the coordinate (\ref{tortoise}) does not belong to the $C^2$-class of admissible coordinate transformations in GR. Indeed, the factor
\begin{equation}
\Psi (r) = \frac {1}{r-r_S}~,
\end{equation}
at the right side of (\ref{tortoise-eq}), represents the discontinuous function across the surface $r = r_S$. If the area of definition of $\Psi (r)$ includes the singular point $r = r_S$, as often is assumed for the tortoise coordinate in (\ref{tortoise}), one needs to introduce the generalized derivative:
\begin{equation}
\frac {d\widehat \Psi}{dr} = \frac {d\Psi}{dr} + [\Psi]_{r=r_S} \delta (r-r_S)~,
\end{equation}
where $\delta (r-r_S)$ is the Dirac delta function and $[\Psi]_{r=r_S}$ denotes the jump of $\Psi$ across $r = r_S$. Since a generalized function that satisfies
\begin{equation} \label{r-psi}
(r-r_S) \psi(r) = 0
\end{equation}
is proportional to Dirac's delta,
\begin{equation} \label{psi=delta}
\psi(r) = C \delta (r-r_S)~,
\end{equation}
where $C$ is some constant, the solution of
\begin{equation} \label{dPsi}
(r-r_S) \frac {d\widehat \Psi}{dr} = 1
\end{equation}
is the function
\begin{equation} \label{Psi=}
\widehat \Psi(r) = \ln |r - r_S| + C_1 + C H(r - r_S)~,
\end{equation}
where $C_1$ is an integration constants and $H(r - r_S)$ is the Heaviside function (see, for example, (3.57) in \cite{Generalize}). We note that, of course one can consider  the solution of (\ref{r-psi}) in ordinary functions and obtain the trivial solution by setting  $C = 0$ in (\ref{psi=delta}). But, in ordinary function the solution (\ref{Psi=}) (when $C = 0$) is valid only if $r \ne r_S$.

So, if we consider (\ref{tortoise-eq}) over the entire interval of definition of the variable $r$ (including the origin $r=0$), the generalized equation,
\begin{equation} \label{tortoise-gen}
(r - r_S) \frac {d\widehat r^*}{dr} = r~,
\end{equation}
should contain the Dirac delta function at the right hand side,
\begin{equation} \label{tortoise-gen'}
\frac {d\widehat r^*}{dr} = \frac {r}{r - r_S} + C \delta (r-r_S)~,
\end{equation}
where $C$ is a constant. Then, the solution of (\ref{tortoise-gen'}), the generalized Regge-Wheeler tortoise coordinate (\ref{tortoise}), incorporates the Heaviside function,
\begin{equation} \label{tortoise'}
r^* =  r + r_S \ln \left( \frac {r}{r_S} - 1\right) + C H(r - r_S)~.
\end{equation}
In the class of generalized functions, singular coordinate transformations of (\ref{Schwarzschild}) (like considered by Kruskal-Szekeres, Eddington-Finkelstein, Lema\^{\i}tre, or Gullstrand-Painlev\'{e}) give $\delta$-functions in the second derivatives, since they all contain one of the factors $\sqrt{r_S - r}$, or $\ln |r_S - r|$. This means that transformed metric tensors at $r=r_S$ are not differentiable, i.e. are of unacceptable class $C^0$ (not of $C^2$, or $C^1$). The vacuum Einstein equation for these metrics is altered with $\delta$-sources at $r=r_S$ that contradicts the validity of the Einstein hole argument.

It is instructive also to recall how the Schwarzschild solution (\ref{Schwarzschild}) was obtained. For the spherically symmetric case the only non-trivial Einstein equation has the form (see, for example \cite{Weinberg:1972kfs}):
\begin{equation} \label{Einstein}
{\cal R}_{rr} = \frac {1}{2r g_{tt}} \,\frac {d {\cal R}_{\theta\theta}}{dr} = 0~,
\end{equation}
where $g_{tt} = - 1/g_{rr}$ is the $tt$-component of the metric tensor and
\begin{equation}
{\cal R}_{\theta\theta} = r \frac {d g_{tt}}{dr} + g_{tt} - 1~.
\end{equation}
It is obvious that the equation (\ref{Einstein}) has two special points, $r = 0$ and $g_{tt} = 0$. In regular points (\ref{Einstein}) is completely equivalent to
\begin{equation} \label{Einstein-2}
\frac {d {\cal R}_{\theta\theta}}{dr} = 0~.
\end{equation}
To find $g_{tt}$ it is sufficient to set ${\cal R}_{\theta\theta}$ to zero, which leads to the equation
\begin{equation} \label{dg}
r \frac {d g_{tt}}{dr} = 1 - g_{tt}~.
\end{equation}
It is clear that this equation in generalized functions, in analogy of (\ref{dPsi}), contains the Heaviside function at $r = 0$. Then, its solution, the Schwarzschild metric (\ref{Schwarzschild}), corresponds to the delta-like source at the origin. In the case of the presence of the tortoise radial variable $r^*$ in (\ref{Einstein-2}) (instead of $r$), after transformation to the Schwarzschild coordinates using (\ref{tortoise-eq}), we find
\begin{equation}
g_{tt} \frac {d {\cal R}_{\theta\theta}}{dr} = 0~,
\end{equation}
what means that the solution contains now the additional delta-function at $g_{tt} = 0$, i.e. on the Schwarzschild horizon. This analysis justifies that (\ref{Schwarzschild}) and (\ref{Schwarzschild-Tortoised}) are solutions to the different equations, corresponding to non-identical sources.

We conclude that The point $r = r_S$ should be excluded from the classical definition of the interval for the Regge-Wheeler tortoise coordinate (\ref{tortoise}), and one needs to set adequate boundary conditions at $r = r_S$ for the equations of matter fields.

%%%%%%%%%%%%%%%%%%%%%%%%%%%%%%%%%%%%%%%%%%%%%%%%%%%%%%%%%%%%%%%%%%%%%%%%%%%%%%%%

\section{Schwarzschild geodesics} \label{Geodesic}

In GR a geodesic $x^\nu (\tau)$ is the extremum of the action integral along the curve,
\begin{equation} \label{S_geo}
S = \int d\tau L~,
\end{equation}
where $\tau$ is an affine parameter. For a particle with nonzero mass $m$, the affine parameter typically taken to be proper time, which mathematically is defined via a metric tensor,
\begin{equation} \label{tau}
d\tau^2 = g_{\mu\nu}dx^\mu dx^\nu~.
\end{equation}
Due to this definition, the 4-velocity $u^\nu = dx^\nu/d\tau$ obeys the relation:
\begin{equation} \label{u^2}
g_{\mu\nu}u^\mu u^\nu = constant~.
\end{equation}

It is natural to believe that the lagrangian of classical particles in (\ref{S_geo}) should be basically non-degenerate. That is, the Hessian matrix of second derivatives of $L(x^\nu, u^\nu)$ with respect to the velocities, $\partial^2 L / \partial  u^\mu \partial  u^\nu$, should be invertible. For singular lagrangians, when the Hessian obeys the condition:
\begin{equation} \label{Hessian}
{\rm det} \left |\frac {\partial^2 L}{\partial  u^\mu \partial  u^\nu} \right | = 0~,
\end{equation}
the system that describes canonical momenta $p^\nu = \partial L/\partial u^\nu$ cannot be solved for the velocities as functions of the coordinates and momenta. This leads to the non-equivalence of the Euler-Lagrange and Hamilton methods. One cannot worry about this, since there exists the Dirac–Bergmann algorithm for converting a theory with a singular Lagrangian into a constrained Hamiltonian system \cite{Salisbury:2006dul, CHS}. The formalism is elegant, but consists of a large number of logical steps and is rather complex.

In the context of Newtonian mechanics, the Lagrangian and the Hamiltonian formulations often are equivalent, since equations of motion in both cases usually describe the same trajectory. However, in field theory the issue of singular lagrangians cannot be avoided. Nearly every physical theory -- electrodynamics, Yang-Mills, GR, relativistic string models -- has gauge freedom and introduces singular lagrangians. However, in these theories lagrangians that come directly from the Principle of Least Action are considered to be 'more fundamental', and Hamiltonians are defined using Legendre transformations.

In particle mechanics one has the freedom to choose the lagrangian among several possibilities, some of which are regular and others singular. Most often the choice is the 'geometrical' lagrangian that gives the arc length of a curve \cite{Pauli}:
\begin{equation} \label{L_1}
L_1 = - m \sqrt{g_{\mu\nu}u^\mu u^\nu}~.
\end{equation}
This lagrangian has nice geometrical interpretation, but in some sense is the ‘worst’ choice to consider motion of a test particle. Indeed, for an arbitrary function of the invariant (\ref{u^2}), the only function that leads to a singular lagrangian is the square root. To show singular features of (\ref{L_1}), consider the radial particle ($u_\theta = u_\phi = 0$) of the unit mass ($m=1$), for which we can construct the two components of momenta,
\begin{equation}
\begin{split} \label{p}
p_t =& \frac{\partial L_1}{\partial u_t} = - \left(1 - \frac {r_S}{r} \right) \frac {u_t}{\sqrt{L_1}}~, \\
p_r =& \frac{\partial L_1}{\partial u_r} = \left(1 - \frac {r_S}{r} \right)^{-1} \frac {u_r}{\sqrt{L_1}}~.
\end{split}
\end{equation}
It is not difficult to see that there exists a relation between them:
\begin{equation}
\left(1 - \frac {r_S}{r} \right)^{-1}p_t^2 = \left(1 - \frac {r_S}{r} \right)p_r^2 + 1~,
\end{equation}
which represents the energy–momentum relation in effective (1+1)-Schwarzschild space for the particle with the unit mass. This constraint is the reason that the geometric lagrangian (\ref{L_1}) appears to be singular, what can be explicitly checked calculating the determinant of the $2 \times 2$ Hessian matrix (\ref{Hessian}) having the components:
\begin{equation}
\frac{\partial^2 L_1}{\partial u_t^2} = -\frac{u_r^2}{\sqrt{L_1}}~,  \qquad \frac{\partial^2 L_1}{\partial u_r^2} = -\frac{u_t^2} {\sqrt{L_1}}~, \qquad
\frac{\partial^2 L_1}{\partial u_t \partial u_r} = - \frac{u_t u_r} {\sqrt{L_1}}~.
\end{equation}

Another popular choice for the lagrangian of relativistic particles in curved space is the expression of the relativistic kinetic energy \cite{Pauli}:
\begin{equation} \label{L_2}
L_2 = \frac m2 g_{\mu\nu}u^\mu u^\nu~.
\end{equation}
This lagrangian is not singular and one can directly construct the Hamiltonian by the familiar procedure:
\begin{equation}
H =  u^\nu \frac{\partial L_2}{\partial u^\nu} - L_2 ~.
\end{equation}

As we have shown, in general, descriptions by the lagrangians (\ref{L_1}) and (\ref{L_2}) are not the same. Both lagrangians give the same equations of motion if they take a constant value for the solution, i.e. two formalisms become equivalent only if we fix the condition (\ref{u^2}). But, for the Schwarzschild case (\ref{Schwarzschild}) this condition is satisfied only if $r \ne r_S$.

Let us consider some problems with singularities for the Schwarzschild geodesics. In literature, a radial motion ($\theta = \pi/2$ and $\phi = 0$) in the Schwarzschild field usually is described not from the second order differential equations for geodesics, but directly from the definition of the proper time (\ref{tau}),
\begin{equation} \label{u^2-Sch}
g_{\mu\nu}u^\mu u^\nu =  \left(1 - \frac {r_S}{r} \right) u_t^2 - \left( 1 - \frac {r_S}{r} \right)^{-1} u_r^2 ~.
\end{equation}
Since $t$ not appears in the metric coefficients, the Euler-Lagrange equations tell us that $p_t = E$ in (\ref{p}) is the constants of motion --  the conserved total energy of a test particle per a unit proper mass $m = 1$,
\begin{equation} \label{dt/dtau}
E =  \left(1 - \frac {r_S}{r} \right) \frac {dt}{d\tau}~.
\end{equation}
Usually the coefficient in this expression is treated as some kind of Lorentz factor and in geodesic equations replacements of $\tau$ with $t$ is made. However, in the ratio of $dt/d\tau$ only $d\tau$ is a function of 4-coordinates, hence, $d\tau = d\tau(t)$ and it seems that the $\tau$ with $t$ are not directly replaceable \cite{MTW}. Using the relation (\ref{dt/dtau}) for photons, $L = 0$, we arrive at the equation
\begin{equation} \label{dr/dtau}
0 = E^2 - \left(\frac {dr}{d\tau}\right)^2~.
\end{equation}
The system (\ref{dt/dtau}) -- (\ref{dr/dtau}) can be solved in quadratures:
\begin{equation} \label{t=}
\tau = const~, \qquad t = \pm \int dr \left( 1 - \frac {r_S}{r} \right)^{-1} = \pm \int dr^* = const \pm r^*~.
\end{equation}
But, according to (\ref{tortoise'}), we need to insert the Heaviside function in the definition of $r^*$ and we find:
\begin{equation} \label{t=r}
t \sim \pm \left[r + r_S \ln \left( \frac {r}{r_S} - 1\right)\right] + C H(r - r_S)~.
\end{equation}

Note that observables should not depend on the choice of formalism. The fact of crossing the given point, like Schwarzschild surface, must be an event having an absolute logical meaning. In the traditional approach there are opposite answers: “yes” from the local observer, and “no” from the distant observer. These predictions follow from suggested formulas rather than adequate solutions of the equations of motion in different coordinate systems. The appearance of the Heaviside function in the expression (\ref{t=r}) means the existence of the real singularity for geodesics at the Schwarzschild radius $r = r_S$ that cannot be removed by coordinate transformations. Then, in equations of particles one needs to set adequate boundary conditions at that point, which eliminates contradictions between results of different observers.

As an example of the importance of boundary conditions at a singularity, consider the ordinary Laplace equation in Cartesian coordinates,
\begin{equation} \label{Laplace}
\left( \partial_x^2 + \partial_y^2 + \partial_z^2 \right){\cal \varphi } = 0 ~.
\end{equation}
If we look for the solution of this equation in regular functions, we obtain the plain waves-type solution of the form:
\begin{equation} \label{plane}
{\cal \varphi } \sim e^{\pm [i (ax + by) + \sqrt {a^2 + b^2}z]}~,
\end{equation}
where $a$ and $b$ are integration constants \cite{Jackson}. However, the most important solution of the Laplace equation (\ref{Laplace}) is the Newtonian potential in spherical coordinates,
\begin{equation} \label{Newton}
{\cal \varphi } \sim \frac 1r ~.
\end{equation}
In generalized functions the singular solution (\ref{Newton}) corresponds to some $\delta$-source at the origin, which should be placed at right hand side of (\ref{Laplace}). This means that the solution (\ref{Newton}) does not follow from the regular boundary conditions at the origin, $r=0$.

Analogously, particle equations on the Schwarzschild background cannot have plane wave solutions on the singular surface $r = r_S$. Nevertheless, based on the solution (\ref{t=}), for infalling particles the boundary conditions usually are imposed assuming the existence of the horizon crossing plane waves \cite{Star, Matz},
\begin{equation} \label{solution-f=0}
\Phi_{r \to r_S} \sim  e^{\pm i (\omega t + k r^*)} ~,
\end{equation}
where the radial variable is represented by the Regge-Wheeler coordinate (\ref{tortoise}). Based on these inadequate boundary conditions and on the assumption that by specific singular coordinate transformations it is possible to remove singularities in the geodesic equations, one can conclude that particles can freely fall through the event horizon \cite{BH-1, BH-2, BH-3}. However, due to the appearance of the $\delta$-function in the second derivatives of $r^*$,  the solution (\ref{solution-f=0}) does not obey the sourceless equation at $r = r_S$.

Alternatively, by applying physical boundary conditions at $r \to r_S$, in \cite{Gogberashvili:2017xti, Gogberashvili:2016xcx} it was found the exponentially decay/enhanced wave functions (with the complex phases $\omega = \pm i/r_S$),
\begin{equation} \label{T}
\Phi_{r \to r_S} \sim  e^{\pm t/r_S} ~,
\end{equation}
which is very different from the familiar internal \cite{BH-wave-1, BH-wave-2} and external \cite{BH-out} periodic-in-time solutions (\ref{solution-f=0}) for the particles close to the Schwarzschild horizon. The existence of exponentially dumping (increasing) in time solutions (\ref{T}) shows that quantum particles do not cross freely the Schwarzschild horizon, but are absorbed or reflected by it.

Note that appearance of complex phases in wavefunctions usually is connected to tunneling or radiation processes. The solutions of wave equations in Schwarzschild's coordinates having the complex phase was obtained also in \cite{Tunneling-1, Tunneling-2, Tunneling-3, Tunneling-4}, were it was nevertheless assumed that classical geodesics are extendable across the horizon, while the real-valued exponential factor was connected with the process of particle creation by the gravitational field of BHs. But, in these papers the singular point at the BH horizon, which shows that classical particles are stopped from entering the Schwarzschild sphere, was removed in propagators by the introduction of the infinitesimal integration contours around the pole.

%%%%%%%%%%%%%%%%%%%%%%%%%%%%%%%%%%%%%%%%%%%%%%%%%%%%%%%%%%%%%%

\section{Conclusion} \label{Concl}

In this article we attempt to show that, even on the level of classical GR, extensions of geodesics across the Schwarzschild horizon by singular diffeomorphisms is problematic. Singular coordinate transformations used to demonstrate continuity of geodesics at a BH horizon, usually are class $C^0$ functions. Singular metric tensors introduce $\delta$-like sources on the surfaces of discontinuity, while, according to Einstein's hole argument, vacuum metric solutions are equivalent only if they correspond to the same energy-momentum tensor in the source region. We explicitly show that the correct expression for the Regge-Wheeler radial variable should contain Heaviside function at the Schwarzschild horizon that corresponds to delta-like sources there. This means that particle equations have not horizon crossing plane wave solutions and we need to set adequate boundary conditions on the Schwarzschild sphere. In this case one obtains the exponential solutions with the complex phases that correspond to the absorptions and reflections of particles by the BH horizon \cite{Gogberashvili:2017xti, Gogberashvili:2016xcx}. In this model the event horizon surrounds an impenetrable sphere of the ultra-dense matter, reflected from which particles could obtain energy from the gravitational field and imitate some burst-type signals, like GRBs, FRBs, ultra-high energy cosmic rays, or the LIGO events \cite{Beradze:2021akh}.

%%%%%%%%%%%%%%%%%%%%%%%%%%%%%%%%%%%%%%%%%%%%%%%%%%%%%%%%%%%%%%%%%%%%%%%%%%%%%%%


\begin{thebibliography}{99}

\bibitem{Oks:2021hef} E.~Oks,
``Brief review of recent advances in understanding dark matter and dark energy,''
New Astron. Rev. \textbf{93} (2021) 101632,
doi: 10.1016/j.newar.2021.101632
[arXiv: 2111.00363 [astro-ph.CO]].

\bibitem{Marolf:2017jkr} D.~Marolf,
``The Black Hole information problem: past, present, and future,''
Rept. Prog. Phys. \textbf{80} (2017) 092001,
doi: 10.1088/1361-6633/aa77cc
[arXiv: 1703.02143 [gr-qc]].

\bibitem{Witten:2019qhl} E.~Witten,
``Light rays, singularities, and all that,''
Rev. Mod. Phys. \textbf{92} (2020) 045004,
doi: 10.1103/RevModPhys.92.045004
[arXiv: 1901.03928 [hep-th]].

\bibitem{Man-San} E.~Minguzzi and M.~Sanchez,
                 {\it The Causal Hierarchy of Spacetimes in Recent Developments in Pseudo-Riemannian Geometry}, ESI Lect. Math. Phys.
                 (Eur. Math. Soc. Publ. House, Z\"{u}rich 2008)
                 [arXiv:gr-qc/0609119 [gr-qc]].

\bibitem{Sorkin:1995ca} R.~D.~Sorkin and E.~Woolgar,
``A causal order for space-times with C0 Lorentzian metrics: Proof of compactness of the space of causal curves,''
Class. Quant. Grav. \textbf{13} (1996) 1971,
doi: 10.1088/0264-9381/13/7/023
[arXiv: gr-qc/9508018 [gr-qc]].

\bibitem{Synge} J.~L.~Synge,
               {\it Relativity: The General Theory}
               (North-Holland Publishing Company, Amsterdam 1960).

\bibitem{hole} R.~Torretti,
{\it Relativity and Geometry}
(Dover, NY 1996).

\bibitem{Act-Pas} D.~J.~ Oliver,
``Validity of the Einstein hole argument,''
Stud. Hist. Philos. Sci. Part B: Stud. Hist. Philos. Mod. Phys.  \textbf{68} (2019) 62,
doi: 10.1016/j.shpsb.2019.04.008
[arXiv: 1907.01614 [physics.hist-ph].

\bibitem{Ansoldi:2008jw} S.~Ansoldi,
``Spherical black holes with regular center: A Review of existing models including a recent realization with Gaussian sources,''
[arXiv: 0802.0330 [gr-qc]].

\bibitem{Cardoso:2019rvt} V.~Cardoso and P.~Pani,
``Testing the nature of dark compact objects: a status report,''
Living Rev. Rel. \textbf{22} (2019) 4,
doi: 10.1007/s41114-019-0020-4
[arXiv: 1904.05363 [gr-qc]].

\bibitem{Modif-1} S.~B.~Giddings,
``Black hole information, unitarity, and nonlocality,''
Phys. Rev. D \textbf{74} (2006) 106005,
doi: 10.1103/PhysRevD.74.106005
[arXiv: hep-th/0605196 [hep-th]].

\bibitem{Modif-3} J.~Maldacena and L.~Susskind,
``Cool horizons for entangled black holes,''
Fortsch. Phys. \textbf{61} (2013) 781,
doi: 10.1002/prop.201300020
[arXiv: 1306.0533 [hep-th]].

\bibitem{Malafarina:2017csn} D.~Malafarina,
``Classical collapse to black holes and quantum bounces: A review,''
Universe \textbf{3} (2017) 48,
doi: 10.3390/universe3020048
[arXiv: 1703.04138 [gr-qc]].

\bibitem{Howard:1984qp} K.~W.~Howard and P.~Candelas,
``Quantum stress tensor in Schwarzschild space-time,''
Phys. Rev. Lett. \textbf{53} (1984) 403,
doi: 10.1103/PhysRevLett.53.403.

\bibitem{tHooft:1984kcu} G.~'t Hooft,
``On the quantum structure of a black hole,''
Nucl. Phys. B \textbf{256} (1985) 727,
doi: 10.1016/0550-3213(85)90418-3.

\bibitem{Mathur:2005zp} S.~D.~Mathur,
``The Fuzzball proposal for black holes: An Elementary review,''
Fortsch. Phys. \textbf{53} (2005) 793,
doi: 10.1002/prop.200410203
[arXiv: hep-th/0502050 [hep-th]].

\bibitem{Modif-2} A.~Almheiri, D.~Marolf, J.~Polchinski and J.~Sully,
``Black holes: Complementarity or firewalls?,''
JHEP \textbf{02} (2013) 062,
doi: 10.1007/JHEP02(2013)062
[arXiv: 1207.3123 [hep-th]].

\bibitem{Einstein} A.~Einstein,
``On a stationary system with spherical symmetry consisting of many gravitating masses,''
Ann. Math. \textbf{40} (1939) 922.

\bibitem{Schwarz} K.~Schwarzschild,
``On the gravitational field of a mass point according to Einstein’s theory,''
Sitzungsber. Preuss. Akad. Wiss., Phys. Math. Kl. (1916) 189.

%%%%%%%%%%%%%%%%%%%%%%%%%%%%%%%%%%%%%%%%%%%%%%%%%%%%%%%%%%%%%%%%%%

\bibitem{Ha-2} Y.~K.~Ha,
``The irreducible mass of Christodoulou-Ruffini-Hawking mass formula,''
Int. J. Mod. Phys. D \textbf{31} (2022) 2230006,
doi: 10.1142/S0218271822300063.

\bibitem{BH-1} S.~Chandrasekhar,
              {\it The Mathematical Theory of Black Holes}
              (Clarendon, New York 1983).

\bibitem{BH-2} S.~Carroll,
              {\it Spacetime and Geometry: An Introduction to General Relativity}
              (Addison-Wesley, San Francisco 2004).

\bibitem{BH-3} E.~Poisson,
              {\it A Relativist's Toolkit: The Mathematics of Black-Hole Mechanics}
              (Cambridge University Press, Cambridge 2004).

\bibitem{Levi-Civita} T.~Levi-Civita,
                     {\it The Absolute Differential Calculus}
                     (Dover, New York 1977).

\bibitem{Crothers} S.~J.~Crothers,
``On the geometry of the general solution for the vacuum field of the point-mass,''
Progress in Phys. \textbf{2} (2005) 3.

\bibitem{Dix} W.~G.~Dixon,
``The definition of multipole moments for extended bodies,''
Gen. Rel. Grav.  \textbf{4} (1973) 199,
doi: 10.1007/BF02412488.

%%%%%%%%%%%%%%%%%%%%%%%%%%%%%%%%%%%%%%%%%%%%%%%%%%%%%%%%%%%%%%%%

\bibitem{Macdonald} A.~Macdonald,
``Einstein’s hole argument,''
 Am. J. Phys. \textbf{69} (2001) 223,
 doi: 10.1119/1.1308265.

%%%%%%%%%%%%%%%%%%%%%%%%%%%%%%%%%%%%%%%%%%%%%%%%%%%%%%%%%%%%%%%%%%

\bibitem{Generalize} F.~Farassat,
                    {\it Introduction to Generalized Functions with Applications in Aerodynamics and Aeroacoustics}
                    (Langley Research Center, Virginia 1994).

\bibitem{Weinberg:1972kfs} S.~Weinberg,
                          {\it Gravitation and Cosmology: Principles and Applications of the General Theory of Relativity}
                          (John Wiley and Sons, NY 1972).

%%%%%%%%%%%%%%%%%%%%%%%%%%%%%%%%%%%%%%%%%%%%%%%%%%%%%%%%%%%%%%%%%%%

\bibitem{Salisbury:2006dul} D.~C.~Salisbury,
``Peter Bergmann and the invention of constrained Hamiltonian dynamics,''
Einstein Stud. \textbf{12} (2012) 247,
doi: 10.1007/978-0-8176-4940-1\_11
[arXiv: physics/0608067 [physics]].

\bibitem{CHS} A.~Hanson, T.~Regge, and C.~Teitelboim,
{\it Constrained Hamiltonian Systems}
(Accademia Nazionale dei Lincei, Roma 1976).

\bibitem{Pauli} W.~Pauli,
               {\it Theory of Relativity}
               (Pergamon Press, NY 1958).

\bibitem{MTW} C.~W.~Misner, K.~S.~Thorne and J.~A.~Wheeler,
             {\it Gravitation}
             (Freedman and Company, San Francisco 1973).

\bibitem{Jackson} J.D. Jackson,
                 {\it Classical Electrodynamics}
                 (Wiley, NY 1999).

\bibitem{Matz} R.~A.~Matzner,
``Scattering of massless scalar waves by a Schwarzschild 'singularity',''
J. Mat. Phys. \textbf{9} (1968) 163,
doi: 10.1063/1.1664470.

\bibitem{Star} A.~A.~Starobinskii,
``Amplification of waves during reflection from a rotating 'black hole',''
Sov. Phys. JETP \textbf{37} (1973) 28.

\bibitem{Gogberashvili:2017xti} M.~Gogberashvili,
``Can Quantum Particles Cross a Horizon?,''
Int. J. Theor. Phys. \textbf{58} (2019) 3711,
doi: 10.1007/s10773-019-04242-0
[arXiv: 1712.02637 [gr-qc]].

\bibitem{Gogberashvili:2016xcx} M.~Gogberashvili and L.~Pantskhava,
``Black hole information problem and wave bursts,''
Int. J. Theor. Phys. \textbf{57} (2018) 1763,
doi: 10.1007/s10773-018-3702-x
[arXiv: 1608.04595 [physics.gen-ph]].

\bibitem{BH-wave-1} T.~Damour and R.~Ruffini,
``Black-hole evaporation in the Klein-Sauter-Heisenberg-Euler formalism,''
Phys. Rev. D \textbf{14} (1976) 332,
doi: 10.1103/PhysRevD.14.332.

\bibitem{BH-wave-2} S.~Sannan,
``Heuristic derivation of the probability distributions of particles emitted by a black hole,''
Gen. Rel. Grav. \textbf{20} (1988) 239,
doi: 10.1007/BF00759183.

\bibitem{BH-out} E.~Elizalde,
``Series solutions for the Klein-Gordon equation in Schwarzschild space-time,''
Phys. Rev. D \textbf{36} (1987) 1269,
doi: 10.1103/PhysRevD.36.1269.

\bibitem{Tunneling-1} K.~Srinivasan and T.~Padmanabhan,
``Particle production and complex path analysis,''
Phys. Rev. D \textbf{60} (1999) 024007.
doi: 10.1103/PhysRevD.60.024007
[arXiv: gr-qc/9812028].

\bibitem{Tunneling-2} E.~T.~Akhmedov, V.~Akhmedova and D.~Singleton,
``Hawking temperature in the tunneling picture,''
Phys. Lett. B \textbf{642} (2006) 124,
doi: 10.1016/j.physletb.2006.09.028
[arXiv: hep-th/0608098].

\bibitem{Tunneling-3} E.~T.~Akhmedov, V.~Akhmedova, T.~Pilling and D.~Singleton,
``Thermal radiation of various gravitational backgrounds,''
Int. J. Mod. Phys. A \textbf{22} (2007) 1705,
doi: 10.1142/S0217751X07036130
[arXiv: hep-th/0605137].

\bibitem{Tunneling-4} E.~T.~Akhmedov, T.~Pilling and D.~Singleton,
``Subtleties in the quasi-classical calculation of Hawking radiation,''
Int. J. Mod. Phys. D \textbf{17} (2008) 2453,
doi: 10.1142/S0218271808013947
[arXiv: 0805.2653 [gr-qc]].

%%%%%%%%%%%%%%%%%%%%%%%%%%%%%%%%%%%%%%%%%%%%%%%%%%%%%%%%%%%%%%%%

\bibitem{Beradze:2021akh} R.~Beradze, M.~Gogberashvili and L.~Pantskhava,
``Reflective black holes,''
Mod. Phys. Lett. A \textbf{36} (2021) 2150200,
doi: 10.1142/S021773232150200X.

%%%%%%%%%%%%%%%%%%%%%%%%%%%%%%%%%%%%%%%%%%%%%%%%%%%%%%%%%%%%%%%%%%%%%%%%%%%%%%%
\end{thebibliography}
\end{document}